\documentstyle[prl,aps,epsf]{revtex}         

\newcommand{\bec}{Bose--Einstein condensate}
\newcommand{\res}{resonance}
\def\be{\begin{equation}}
\def\ee{\end{equation}}
\def\bea{\begin{eqnarray}}
\def\eea{\end{eqnarray}}

\begin{document}
%
\twocolumn[\hsize\textwidth\columnwidth\hsize\csname @twocolumnfalse\endcsname 

\title{Bragg spectroscopy of a \bec }
\author{J. Stenger, S. Inouye, A.P. Chikkatur, D.M. Stamper--Kurn, D.E. Pritchard,
and W. Ketterle}
\address{Department of Physics and Research Laboratory of
Electronics, \\
Massachusetts Institute of Technology, Cambridge, MA 02139}
\maketitle

\vspace{-2mm}
\begin{center}
\today
\end{center}
\vspace{-2mm}

\begin{abstract}
Properties of a \bec\ were studied by stimulated, two--photon Bragg scattering. The high 
momentum and energy resolution of this method allowed a spectroscopic measurement of the 
mean-field energy and of the intrinsic momentum uncertainty of the condensate. The coherence 
length of the condensate was shown to be equal to its size.  Bragg spectroscopy can be used to 
determine the dynamic structure factor over a wide range of energy and momentum transfers.
\end{abstract}
\vskip1pc
]

%
The first evidence for Bose-Einstein condensation 
in dilute gases was obtained by a sudden narrowing of the 
velocity distribution as observed for ballistically expanding clouds of atoms \cite{ande:95}.  
Indeed, most textbooks describe Bose-Einstein condensation as condensation in momentum space 
\cite{huan:87}. 
However, the dominant contribution to the observed momentum distribution of the 
expanding condensate was the released interaction energy 
(mean--field energy) resulting in momentum 
distributions much \textit{broader} than the zero--point motion of the ground state of the 
harmonic trapping potential.  Since the size of a trapped condensate with repulsive interactions is 
larger than the trap ground state, the momentum distribution should be considerably 
\textit{narrower} than in the trap ground state.  In this paper, we show the momentum 
distribution of a trapped condensate to be Heisenberg 
uncertainty limited by its finite size. This is equivalent to showing 
that the coherence length of the condensate is equal to its physical size.

Sub--recoil momentum resolution has been previously achieved by resolving the Doppler 
width of a Raman 
transition between different hyperfine states \cite{kase:92} or of a two--photon transition to a 
metastable excited state \cite{kill:98}.  Here we use Bragg scattering where two momentum 
states of the \textit{same} ground state are connected by a stimulated two--photon process 
\cite{kozi:98}. This process can be used to probe density fluctuations of the 
system and thus to measure directly the 
dynamic structure factor S({\bf q,$\nu$}), which is the Fourier transform of the 
density--density 
correlation function and is central to the theoretical description of many--body systems 
\cite{grey:78}. 
In contrast to measuring S({\bf q,$\nu$}) with inelastic neutron scattering like
in superfluid helium \cite{soko:95}, or using inelastic light scattering \cite{park:98,poli:96}, 
Bragg scattering as used here is a stimulated process which greatly enhances 
resolution and sensitivity.

Bragg scattering of atoms from a light grating
was first demonstrated in 1988 \cite{mart:88a} and has been 
used to manipulate atomic samples in atom interferometers \cite{gilt:95}, in 
\mbox{de Broglie} wave frequency shifters \cite{bern:96}, and also to couple out or manipulate 
a Bose--Einstein condensate \cite{kozi:98}.  
Small angle Bragg scattering, called recoil--induced \res s, has been used for thermometry
of laser--cooled atoms \cite{cour:94}.
In this work we establish Bragg 
scattering as a spectroscopic technique to probe properties of the condensate. We refer to it as 
Bragg spectroscopy in analogy to Raman spectroscopy which involves different \textit{internal}
states.

The absorption of $N$ photons from one laser beam and stimulated emission into a second laser 
beam constitutes an $N$--th order Bragg scattering process. The momentum transfer $q$ and 
energy transfer $h\nu$ are given by
$q = 2N\hbar k \sin(\vartheta /2)$ and $\nu = N \Delta \nu$,
where $\vartheta$ is the angle between the two laser beams with wave vector $k$ and 
frequency difference $\Delta \nu.$

For non--interacting atoms with initial momentum 
$\hbar k_i$, the resonance is given by the Bragg condition
$
h\nu =  q^2/2m + \hbar k_iq/m 
$, 
which simply reflects energy and momentum conservation for a free particle.
The second term is the Doppler shift of the 
resonance and allows the use of Bragg resonances in a velocity--selective way
\cite{kozi:98,cour:94}.

For a weakly interacting homogeneous condensate at density $n$, the dispersion relation has the 
Bogoliubov form \cite{huan:87}

\be
\nu = \sqrt{\nu_0^2 + 2\nu_0 n U/h}  \label{mfshift} ,
\ee
where $nU = n4\pi\hbar^2a/m$ is the chemical potential, with $a$ and $m$ 
denoting the scattering length and the mass, respectively, and $h\nu_0 = q^2/2m$.
At low energies the excitation spectrum is phonon--like,
obeying $h\nu= c q$, where $c$ is the speed of sound \cite{huan:87}. 
For energies $h\nu \gg nU$ the spectrum is particle--like 

\be 
\nu \approx \nu_0 + n U/h .\label{mfshiftapprox}
\ee
The mean--field shift $nU$
reflects the exchange term in the interatomic interactions: a particle 
with momentum $q$ experiences \textit{twice} the mean field energy as a particle in the 
condensate \cite{huan:87}. 
We use this property to determine the condensate mean--field energy spectroscopically.
This is related to, but different from the mean--field shift due to interactions with an 
electronically excited state which was used to identify BEC in atomic hydrogen \cite{kill:98}.

Eq. (\ref{mfshift}) is the excitation spectrum of a homogeneous condensate with 
initial momentum $\hbar k_i=0$.  The inhomogeneous trapping potential adds 
two features which broaden the resonance: a distribution 
of initial momenta due to the finite size of the cloud, and a distribution of mean--field shifts 
due the density variation over space.

The momentum distribution along the x--axis is given by the 
Fourier transform of the wavefunction:
$
f(p_x) = \left[ \int dx\, dy\, dz\, e^{-ip_x x /\hbar} \Psi (x,y,z) \right]^2
$.
In the Thomas--Fermi approximation the wave function $\Psi (x,y,z)$ in an harmonic 
trapping potential is $\left[ \Psi (x,y,z) \right]^2 = n_0(1-(x/x_0)^2-(y/y_0)^2-(z/z_0)^2)$, where
$n_0$ denotes the peak density.
The size of the wavefunction is determined by $n_0$ and the trapping frequencies $\nu_i$ 
through the coefficients $x_0, y_0, z_0$: $x_0=\sqrt{2n_0U/m(2\pi\nu_x)^2}$ (and similar for 
$y_{0}, z_{0}$). The line shape $I_p(p_{x})$ is proportional to the 
square of the Fourier coefficients \cite{baym:96}

\be 
I_p(p_x) \sim \left[ J_2(p_x x_0 /\hbar)/(p_x x_0 /\hbar)^2 \right]^2 ,  \label{shapefs}
\ee
where $J_2$ denotes the Bessel function of order 2.
This curve is very similar to a Gaussian and has an rms--width  
of $\Delta p_x = \sqrt{21/8} \;\hbar/ x_0$. Thus, 
the corresponding Doppler broadening $\Delta \nu_p = \sqrt{21/8}\;q/2\pi m x_0$ 
of the Bragg resonance 
is inversely proportional to the condensate size $x_0$ and does not depend explicitly on the
number of atoms.

The same parabolic wavefunction gives the (normalized) density distribution
$(15n/4n_0) \sqrt{1- n/n_{0}}$. 
The simplest model for the spectroscopic lineshape $I_n(\nu)$ due to 
the inhomogenous density assumes that a volume element with local density $n$ leads to a 
lineshift $nU$ (eq. (\ref{mfshiftapprox})):

\be
I_n(\nu) = 
	\frac{15h(\nu-\nu_0)}{4n_0U} \sqrt{1- \frac{h(\nu -\nu_0)}{n_0U}}, \label{shapemf}
\ee
which has a total width of $n_{0}U/h$, a maximum at $2 n_{0}U/3h$, an average value 
(first moment) of $4 n_{0}U/7h$, and an rms--width of $\Delta\nu_n=\sqrt{8/147} n_{0}U/h$.
In contrast to the Doppler broadening due to the finite size, 
the mean--field broadening only depends on the density but not explicitly on the size.

In our experiments the combined broadening mechanisms represented by 
eqs. (\ref{shapefs}) and (\ref{shapemf}) have to be considered.
While the exact calculation of the lineshape requires detailed knowledge of the 
excitation wavefunctions, the total line shift (first moment) and rms--width can 
be calculated using sum rules and Fermi's Golden Rule. Thus, it can be rigorously shown
that the total line shift remains $4 n_{0}U/7h$, while
the rms--widths $\Delta\nu = \sqrt{\Delta\nu_p^2+\Delta\nu_n^2}$ is the quadrature sum
of the Doppler and mean--field widths \cite{stam:99}.

We produced magnetically trapped, 
cigar--shaped \bec s as in previous work \cite{mewe:96}. In order
to study the \res\ as a function of density and size, we prepared condensates using two different
trapping frequencies and varied the number of atoms by removing a 
variable fraction using the rf--output coupler \cite{mewe:97}.
The density of the condensate was determined from the expansion of the cloud in 
time--of--flight and the size from the atom number and the 
trapping frequencies \cite{mewe:96}.
Bragg scattering was performed by using two counterpropagating beams aligned 
perpendicularly to the weak axis of the trap.  Spectra were taken by pulsing on the light 
shortly before switching off the trap 
and determining the number of scattered atoms as a function of the frequency difference 
between the two Bragg beams. Since the kinetic energy of 
the scattered atoms was much larger than the mean--field energy, they were 
well separated from the unscattered cloud after a typical ballistic expansion time of 20 msec. 
Center frequencies and widths were determined from Gaussian fits to the spectra.

Duration, intensity and detuning of the Bragg pulses had to be chosen carefully. 
The instrumental resolution is limited by 
the pulse duration $\delta t_{pulse}$ due to its Fourier spectrum, in our case requiring
$\delta t_{pulse} > 250 \mu$s for sub--kHz resolution.
The maximum pulse duration was limited to less than one quarter of the trap period,
$\delta t_{pulse} < 500 \mu$s, by which time the initially scattered atoms would come to
rest and thus would be indistinguishable from the unscattered atoms in time--of--flight.
The light intensity was adjusted to set the peak efficiency to about 20 \%. 
Sufficient detuning was necessary to avoid heating of the sample. The ratio of the 
two--photon rate $\omega_R^2/4\Delta$ to the spontaneous scattering rate 
$\omega_R^2\Gamma/2\Delta^2$ is $\Gamma /2\Delta$, where $\omega_R$ 
denotes the single beam 
Rabi frequency, $\Delta$ the detuning and $\Gamma$ the natural linewidth.  Spontaneous 
scattering was negligible for the chosen detuning of 1.77 GHz below the 3S$_{1/2}$ F=1 
$\rightarrow$ 3P$_{3/2}$ F=2 transition.

The relative detuning of the two
Bragg beams was realized in two ways.  In one scheme, a beam 
was split and sent through two independent acousto--optical modulators driven with the 
appropriate difference frequency, and then overlapped in a counterpropagating configuration.
Alternatively, a single beam was modulated with two frequencies separated by the relative 
detuning and backreflected.  Both methods are insensitive to frequency drifts of the laser 
since the Bragg process only depends on the relative frequency of the two beams, which is 
controlled by rf--synthesizers.  
The second method simultaneously scattered atoms in the $+x$ and $-x$ directions and
was thus helpful to identify motion of the cloud.
We estimate that residual vibrational noise 
broadened the line by less than 1 kHz. This resolution corresponds to a velocity 
resolution of 0.3 mm/s or 1 \% of the single--photon recoil.
At a radial trapping frequency of 200 Hz, we had to avoid any motion of the 
cloud with an amplitude larger than 0.2 $\mu m$.

Fig.  1 shows typical spectra, taken both for a trapped condensate and after 3 ms time of 
flight when the mean--field energy was fully converted into kinetic energy.  The rms--width of the 
resonance for the ballistically expanding cloud is 20 kHz, which is much narrower than 
the 65 kHz wide distribution of a thermal cloud at 1 $\mu$K, a typical value for the BEC 
transition temperature under our conditions. 
\begin{figure}[htbf]
\epsfxsize=80mm					
 \centerline{\epsfbox{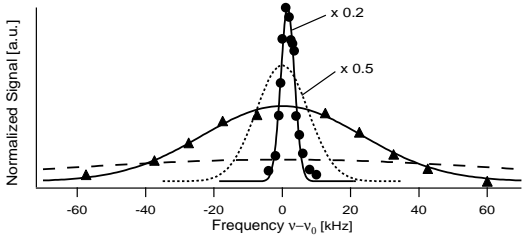}}
\caption{Bragg resonances for a trapped condensate (circles) and after 3 ms time of flight 
(triangles).  For comparison, the momentum distributions of the ground state of the trapping
potential (dots) and of a 1 $\mu$K cold, thermal cloud (dashes) are indicated.}
\end{figure}
We could not measure the thermal distribution 
with the same pulse duration as for the condensate since the fraction of scattered atoms was 
too small due to the broad resonance. The spectra for the thermal cloud and the 
expanding condensate correspond to the spatial distributions observed by absorption imaging 
after sufficiently long time of flight. With this technique, the BEC transition is indicated 
by a sudden narrowing of the time--of--flight distribution by a factor of three. 
Using Bragg spectroscopy, the 
signature of BEC is much more dramatic --- the condensate resonance is more than thirty times 
narrower than of the thermal cloud, and indeed narrower than the ground state of the 
trap. This demonstrates that Bragg spectroscopy is a superior way to probe the BEC 
transition.

\begin{figure}[htbf]
\epsfxsize=80mm					
 \centerline{\epsfbox{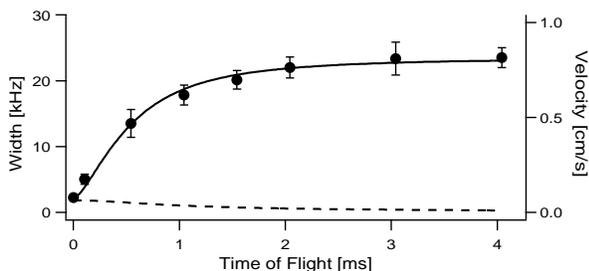}}
\caption{Mean--field acceleration of a condensate released from the trap.
Shown is the increase of the width of the Bragg \res\ during the expansion. The
solid line is the theoretical prediction [18] using the trap frequency $\nu_r=$195 Hz.
The dashed line represents the
contributions of mean--field energy and finite size to the total width.}
\end{figure}

Fig.  2 shows the conversion of mean--field energy into kinetic energy after the atoms are
released from the trap, indicated by a 
broadening of the Bragg resonance from about 2 kHz to 20 kHz.  After 3 ms, the cloud has 
reached its asymptotic velocity $v_\infty$.  
Fig.  2 agrees with the velocity evolution
$v_r = v_\infty 2\pi\nu_r t/\sqrt{1+ (2\pi\nu_r t)^2}$
as expected from the scaling laws given in ref. \cite{cast:96} which are derived 
for cigar shaped condensates with large aspect ratios in the Thomas--Fermi approximation.
In order to compare with the data, the Doppler width due to the velocity evolution, the 
mean--field width and the
finite--size width were added in quadrature, assuming Gaussian shapes. 
The finite--size width was calculated from the predicted evolution of the 
size $x_r = x_0\sqrt{1+ (2\pi\nu_r t)^2}$ \cite{cast:96}.

The narrow resonance of the trapped condensate (Fig. 1) was studied as a function of the 
condensate density 
and size.  Fig. 3 (a) demonstrates the linear dependence of the frequency shift on the 
density.  The slope of the linear fit corresponds to $(0.54 \pm 0.07)\, n_0U/h$, in 
excellent agreement with the prediction of $4n_0U/7h$. In Fig. 3 (b), 
the expected widths due to the mean--field energy 
and finite size are shown for the two 
different trapping frequencies studied.  The data agree well with the solid lines, which 
represent the quadrature sum of the two contributions. To demonstrate the 
finite--size effect the same data are shown in Fig. 3 (c) after subtracting 
the mean--field broadening and the finite pulse--length broadening (0.5 kHz).  
The linewidths are consistent with the expected $1/x_0$ dependence.  
Even without these corrections the 
measured linewidths are within 20 \% of the value expected due to the Heisenberg--uncertainty 
limited momentum distribution  (Fig. 3 (b)). 

The momentum spread of the condensate is limited by its 
coherence length $x_{c}$ which, in the case of long--range order, should be equal to the size 
$x_{0}$ of the condensate. Our results show
that $x_{c} \approx x_{0}$ in the radial direction
of the trap. This quantitatively confirms the earlier
qualitative conclusion reached by interfering two condensates 
\cite{andr:97}. In particular, our measurements indicate that the condensate does not have 
phase fluctuations, i.e. that it does not 
consist of smaller quasi--condensates with random relative phases. It would be 
interesting to study this aspect during the formation of the condensate, e.g. after suddenly 
quenching the system below the BEC transition temperature \cite{mies:98}, 
and observe the disappearance of phase fluctuations.

In this work we have determined the normal mode of the condensate at a momentum transfer of 
two photon recoils, corresponding to a energy transfer $\nu_0=100$ kHz. 
Different momentum transfers are possible 
by changing the angle between the Bragg beams and/or the order $N$, thus enabling 
measurements of the dynamical 
structure function $ S({\bf q},\nu )$ over a wide range of parameters. At low 
momentum transfer, the lineshape is dominated by the mean--field energy and by phonon--like
collective excitations,
whereas at high momentum transfers, the linewidth mainly reflects the 
momentum distribution of individual atoms. This is analogous to neutron scattering in liquid 
helium, where slow neutrons were used to observe the phonon and roton part of the dispersion 
curve, and fast neutrons were used to determine the zero--momentum peak of the condensate 
\cite{soko:95}. 
While we have observed 
higher--order Bragg scattering up to third order in the trapped condensate using higher laser
intensities, its spectroscopic use was precluded by severe Rayleigh scattering, and
would require larger detuning from the atomic resonance.

\begin{figure}[t]
\epsfxsize=75mm					
 \centerline{\epsfbox{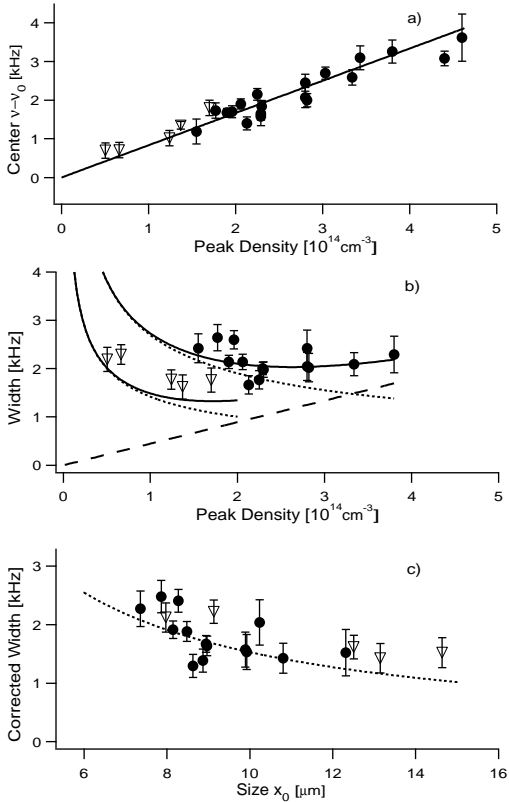}}
\caption{
Bragg spectroscopy of a trapped condensate. Line shifts and widths are shown for
various densities and sizes of the condensate using two different radial trapping frequencies,
$\nu_r = (195\pm 20)$ Hz (circles), and $\nu_r = (95\pm 20)$ Hz (triangles).
The lines in b) show the contributions 
of the mean--field energy (dashed), due to the finite size (dotted, 
for both trapping frequencies), and their quadrature sum (solid lines).
Fig c) displays the width after subtraction of the contributions of the mean--field and the 
finite pulse duration and compares it with the prediction for the momentum uncertainty 
due to the finite size.
The error bars are 1$\sigma$ errors of the Gaussian fits to the data.
}
\end{figure}

The use of inelastic light scattering to determine the structure factor of a \bec\ was discussed 
in several theoretical papers \cite{park:98,poli:96}. It would require the 
analysis of scattered light with kHz resolution and suffers from a strong background of 
coherently scattered light \cite{poli:96}.
Bragg spectroscopy has distinct advantages because it is a stimulated, 
background--free process 
in which the momentum transfer and energy are pre--determined by the laser beams rather than 
post--determined from measurements of the momentum and energy of the scattered particle.  

In conclusion, we have established Bragg spectroscopy as a new tool to measure properties of a 
condensate with spectroscopic precision. We have demonstrated its capability to perform 
high--resolution velocimetry by
resolving the narrow momentum distribution of a trapped condensate and 
by observing the acceleration phase in ballistic expansion.
Since the momentum transfer can be adjusted over a 
wide range, Bragg spectroscopy 
can be used to probe such diverse properties as collective excitations, 
mean--field energies, coherence properties, vortices or 
persistent currents. 

This work was supported by the Office of Naval Research, NSF, Joint Services Electronics 
Program (ARO), NASA, and the David and Lucile Packard Foundation.  A.P.C would like to 
acknowledge support from NSF, D.M.S.-K. from
the JSEP Graduate Fellowship Program and J.S. from the Alexander von 
Humboldt--Foundation.

\bibliographystyle{prsty}

\end{document}